# Directed Self-Assembly of Epitaxial CoFe$_2$O$_4$-BiFeO$_3$ Multiferroic Nanocomposites

*Ryan Comes[1], Hongxue Liu[1], Mikhail Khokhlov[1,2], Richard Kasica[3], Jiwei Lu[1], and Stuart A. Wolf[1,4]*

[1]University of Virginia, Department of Materials Science and Engineering, Charlottesville, VA 22904

[2]Guilford College, Greensboro, NC 27410

[3]National Institute of Standards and Technology, Center for Nanoscale Science and Technology, Gaithersburg, MD 20899

[4]University of Virginia, Department of Physics, Charlottesville, VA 22904

ABSTRACT. CoFe$_2$O$_4$ (CFO)-BiFeO$_3$ (BFO) nanocomposites are an intriguing option for future memory and logic technologies due to the magnetoelctric properties of the system. However, these nanocomposites form with CFO pillars randomly located within a BFO matrix, making implementation in devices difficult. To overcome this, we present a technique to produce patterned nanocomposites through self-assembly. CFO islands are patterned on Nb-doped SrTiO$_3$ to direct the self-assembly of epitaxial CFO-BFO nanocomposites, producing square arrays of CFO pillars.

Multiferroic nanocomposite films have been heavily studied for their potential applications in magnetoelectric systems.[1] The $CoFe_2O_4$-$BiFeO_3$ (CFO and BFO, respectively) system has generated particular interest due to the magnetoelastic properties of CFO[2] and the combination of ferroelectricity and anti-ferromagnetism in BFO[3]. It has been shown that when CFO and BFO are codeposited via physical vapor deposition at high temperatures on a $SrTiO_3$ (001) substrate that the materials will spontaneously phase segregate to produce an epitaxial CFO pillar in an epitaxial BFO matrix, which is referred to as a 1-3 nanocomposite.[4] The CFO pillars form faceted structures with {110}-type interfaces with the BFO matrix and {111}-facets on the surface, protruding above the matrix.[5] The pattern of the CFO pillars in the structure is essentially random, since they are formed through the nucleation of a CFO island on the substrate, while BFO wets the remaining surface. Thus, to control the location of the pillars a means of controlling the nucleation site for the CFO island is needed. CFO-BFO composites have been found to demonstrate magnetoelectric coupling, allowing for electrical control of the magnetic anisotropy of the CFO pillars.[6,7] Based on these properties, the composite system has been proposed for both magnetoelectric memory[8] and logic[9] applications. In particular, the reconfigurable array of magnetic automata (RAMA)[9,10] is a nanomagnetic logic system based on the magnetic quantum cellular automata (MQCA) logic architecture[11] which would use a CFO-BFO 1-3 composite with the pillars arranged in a square array to create a reprogrammable logic system. However, in order to make devices using these composites, the ability to place the pillars into pre-determined arrays is required.

Previous work in patterning multiferroic nanocomposites has been limited. One method to produce patterned magnetoelectric composites is to use a porous anodic aluminum oxide (AAO) film as a liftoff mask during deposition, which produces a hexagonal array pattern.[12,13] In one approach, a $BaTiO_3$-$CoFe_2O_4$ (BTO-CFO) multilayer is deposited onto the AAO film on an STO substrate, which yields a small amount of magnetoelectric response.[12] Another technique is to use the AAO film to form CFO islands and then overcoat the islands with ferroelectric $Pb(Zr,Ti)O_3$ (PZT), which yields a composite that is both ferroelectric and ferromagnetic.[13] Others have used a SiN membrane as a shadow mask to grow ferromagnetic $La_{0.7}Sr_{0.3}MnO_3$ islands and overcoat them with ferroelectric $PbTiO_3$, which produces an epitaxial composite structure with sub-micron dots and intriguing ferroelectric domain structures.[14] However, none of these techniques offers the degree of magnetoelectric control found in the BFO-CFO 1-3 epitaxial nanocomposites.[6,8] Additionally, the AAO and membrane masks are not practical for the formation of a square array of pillars needed for the proposed memory and logic architectures. Another approach involves the use of block copolymers to order the formation of polycrystalline CFO pillars in a polycrystalline PZT matrix using a sol-gel process.[15] This technique also produces a hexagonal array of pillars, but demonstrates a stronger magnetoelectric response than found in the other works. Larger, micron-scale polycrystalline CFO islands embedded in a $PbTiO_3$ matrix have been fabricated using e-beam lithography via a liftoff process, but such techniques are more difficult for smaller nanoscale islands and may not produce the epitaxial pillars and matrix that are desired.[16] A bottom-up technique to arbitrarily define the location of individual pillars in the composite would be ideal for future technologies.

Templated self-assembly is a popular technique to control the formation of a self-assembled structure.[17,18] In this approach, the surface of a substrate is modified to constrain how a pattern will form during thin film deposition. Focused ion beam (FIB) patterning of a Si substrate to form pits has been shown to be effective in creating a preferred nucleation site for the formation of Ge epitaxial quantum dots.[19,20] Similarly, e-beam lithography (EBL) has been used to form SiC on the substrate surface, which acts as a nucleation site during Ge quantum dot growth.[21] However, no reports of templating the growth of oxide composites have been found. In this work, we demonstrate the growth of self-assembled BFO-CFO 1-3 nanocomposites with the pillars patterned into a square array through the use of a substrate with CFO islands on the surface patterned using EBL.

A top-down lithographic process is employed to pattern CFO islands on the substrate surface, followed by a bottom-up self-assembly process to produce the ordered CFO-BFO nanocomposite. A full schematic of the fabrication process is shown in Fig. 1. A 0.5%-Nb-doped STO (Nb:STO) (001) conductive substrate was prepared using common etching and annealing techniques to produce a $TiO_2$ terminated surface with step-edges due to substrate miscut.[22] A 12-nm CFO film was then deposited on the substrate using pulsed electron deposition (PED).[23,24] The CFO film grows via the Volmer-Weber epitaxial growth mode, producing epitaxial islands on the surface.[25] An atomic force microscopy (AFM) surface topography scan after the growth is shown in Fig. 2(a). The sample was found to have islands uniformly coating the surface with diameters between 25 and 50 nm.

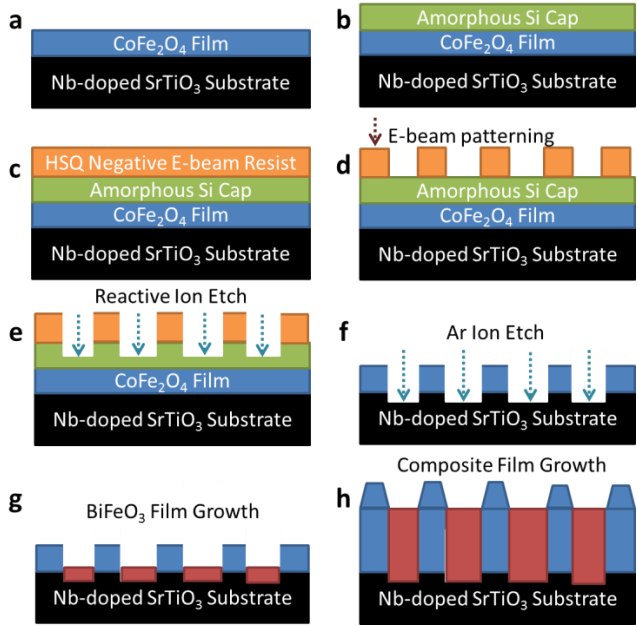

**Figure 1.** Flow chart for the fabrication process. (a) Deposition of $CoFe_2O_4$ (CFO) film using pulsed electron deposition (PED); (b) Deposition of amorphous Si capping layer using RF sputtering; (c) Spin coating of sample with HSQ negative-tone e-beam resist; (d) Patterning of pillars using e-beam lithography; (e) Reactive ion etching of Si cap; (f) Ar ion etching of CFO film; (g) Deposition of 1 nm thick $BiFeO_3$ (BFO) film using PED; (h) Co-deposition of CFO and BFO using PED to form an epitaxial nanocomposite.

An amorphous 20 nm Si capping layer was then sputtered onto the surface of the CFO film in a cleanroom environment. The Si capping layer has previously been shown to be useful in promoting adhesion of hydrosilsesquioxane (HSQ) negative-tone e-beam resist and also acts as a sacrificial etch mask later in the process.[26] HSQ resist was spin coated onto the Si capping layer and arrays of dots 50 nm in diameter with center-to-center distances (pitch) of 100, 150 and 200 nm were then written onto a single substrate, which can be seen in Fig. 2(b). A scanning electron microscope (SEM) image of the patterned array with 200 nm center-to-center spacing (pitch) is shown in Fig. 2(c). The pattern was then transferred into the Si capping layer using a reactive ion etch (RIE) with a mixture of $SF_6$ and $O_2$ gases. Finally, the remaining pattern of HSQ and Si pillars was used as a sacrificial mask for an Ar ion etch using the RIE system. An AFM scan of the islands following the etching step is shown in Fig. 2(d), along with cross-sectional height data in Fig. 2(e). The CFO islands are between 3 and 4 nm tall. It is interesting to note that following the etching process, step-edges from the original substrate annealing process are again visible, indicating a high quality STO surface for the growth of the composite film.

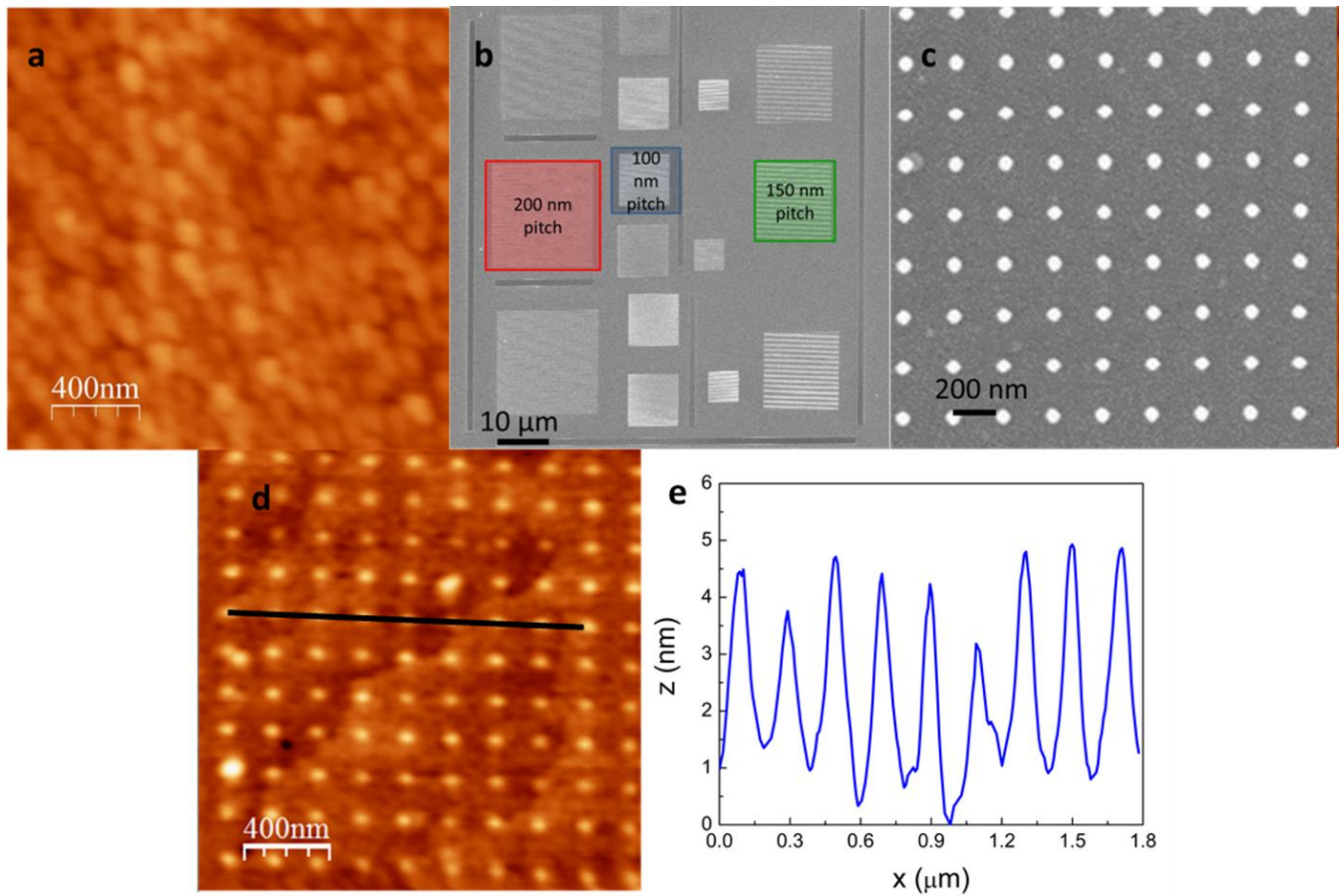

**Figure 2.** (a) Atomic force microscope (AFM) image of uniform CFO film used for template pattern showing regular epitaxial islands on the surface; (b) Scanning electron microscope (SEM) image showing the configuration of arrays on a single sample after EBL; (c) SEM image of EBL-patterned 200

nm pitch array before etching; (d) AFM image of CFO islands after reactive ion etching (RIE) step, showing uniform 200 nm spacing and substrate step edges which are preserved; (e) AFM cross section along black line in (d).

The patterned substrate was then used as a template for a second film grown using the PED system. Details of the deposition conditions are available in the online Supporting Information. An initial 1 nm BFO film was grown by ablating only the BFO target with 20 at. % excess bismuth, which wets the surface and prevents formation of CFO pillars away from the patterned region. BFO is believed to preferentially grow on the substrate surface and leave the CFO pillars relatively unaffected due to the lattice mismatch and immiscibility of the perovskite and spinel phases. Immediately after the conclusion of the initial BFO deposition, the CFO gun was activated and a codeposition was performed, producing an overall BFO thickness of approximately 25 nm based on x-ray reflectivity measurements.

The resulting film was characterized using an Asylum Research Cypher AFM system. AFM analysis was performed on the arrays with pillar pitches of 100, 150 and 200 nm. Images for all three dimensions are shown in Fig. 3, along with cross-sectional height data for each image. The 100 nm and 150 nm pitch arrays have uniformly distributed pillars with no interstitial pillars present. In the 200 nm arrays some small CFO pillars nucleated in spite of the initial BFO wetting layer, an example of which is highlighted with a diamond in Fig. 3(c). A key issue is whether the co-deposited CFO preferentially and completely segregates to the pillar sites.  To address this, detailed statistical analysis of the AFM measurements for each of the three pitch sizes is shown in Table 1. The average pillar height and width above the surface are shown for each of the pitch sizes, which allows for the calculation of the volume of CFO phase held above the surface of the BFO matrix at the template sites for each array. The calculated volume per unit area for all three pitch sizes is shown in Table 1, with the values normalized such that the value for the 100 nm array is equal to 1. Based on this data, we are able to show that the 100 nm and 150 nm arrays have similar CFO volume per unit area (1 and 1.02, respectively) above the surface. This means that CFO mass is conserved above the surface of the matrix. The pillar area coverage for the 100 nm and 150 nm arrays is also similar. Since the pillars are not expected to deviate substantially in diameter during the film growth, this is an indication that the volume of CFO below the surface of the BFO matrix is equal as well. This result agrees with the work of Zheng, *et al,* (Ref. 4), which showed that the area coverage of the pillars beneath the matrix surface is equal to the volume fraction of CFO in the adatom flux.[4] Thus, all CFO flux deposited in the 100 and 150 nm pitch arrays segregates to the pillars at the template sites.

For the 200 nm pitch array, there is excess CFO volume above the surface, as shown by the 1.10 normalized volume per unit area in Table 1. A close inspection of Fig. 3(c) shows that the pillars are more irregularly shaped than in the two arrays with smaller pitch, which indicates that the pillars are most likely not forming the ideal faceted interface with the BFO matrix that is seen in the other arrays. The fractional area coverage of the pillars at the template sites in array is also substantially reduced, meaning that the pillar volume is not sufficient to account for the amount of CFO deposited in the region. The interstitial CFO pillars which form within the BFO matrix in the 200 nm pitch array account for the remainder of the CFO mass within the array.

An analysis of the surface diffusion length for the CFO adatoms is useful to elucidate the kinetics of the growth process. It is clear from the AFM analysis of both the 100 nm and 150 nm pitch arrays that all CFO flux is captured at the template sites. However, this is not the case in the 200 nm pitch array. It would appear that the diffusion length of the CFO adatoms is not sufficient for flux that lands far from any of the template sites to move to one of the sites. The maximum diffusion length, $L$, required for adatoms landing in the BFO matrix to reach a template site is given by:

$$L = \frac{\sqrt{2}}{2} P - r \qquad (1),$$

where $P$ is the array pitch and $r$ is the radius of the pillar. This maximum distance occurs for adatoms that land equidistant from all 4 neighboring template sites at the center of the square formed by the sites. For the 150 nm pitch with a pillar radius of 27 nm, $L$ is equal to approximately 79 nm, and for the 200 nm pitch with a 30 nm diameter, $L$ is equal to 111 nm. Thus we would expect the diffusion length to fall between 79 and 111 nm. The surface diffusion length, $d$, of CFO adatoms can be determined using the formula:

$$d = \sqrt{4Dt} \qquad (2),$$

where $D$ is the surface diffusivity and $t$ is the deposition time to form a unit cell monolayer of BFO in the matrix, which is inversely proportional to the deposition rate. The diffusivity, $D$, is a temperature dependent parameter given by:

$$D = D_0 e^{-E_a/kT} \qquad (3),$$

where $k$ is Boltzmann's constant, $T$ is the surface temperature of the sample, $E_a$ is the activation energy for surface diffusion and $D_0$ is a constant which depends in part on the diffusion geometry but is nominally independent of temperature. Zheng *et al.* (Ref. 4) previously performed calculations to determine the activation energy for surface diffusion, $E_a$, of CFO adatoms in the BFO matrix and found a value of 1.66 eV with no stated uncertainty. By further analyzing the results presented in that work we can perform a linear fit to estimate the value for $D_0$, and find a value of $1.34 \pm 0.91 \times 10^{10}$ nm$^2$/sec.[4] For further details on the fitting calculation and assumptions involved, see the online Supporting Information. With the value of $D_0$ determined and the published value of $E_a$ with a small uncertainty of 0.05 eV assumed, it becomes possible to estimate the diffusion length of the CFO adatom flux. We find a value for $d$ equal to $55 \pm 26$ nm, which is in reasonable agreement with the experimental results since $L$ = 79 nm is within the uncertainty of the calculation for $d$. This shows that interstitial pillars only form in the 200 nm pitch array due to the kinetic limitations during the growth. In order to produce a high quality array with pillar spacing of 200 nm, a slower deposition rate or higher substrate temperature would be required, which could increase the adatom diffusion length.

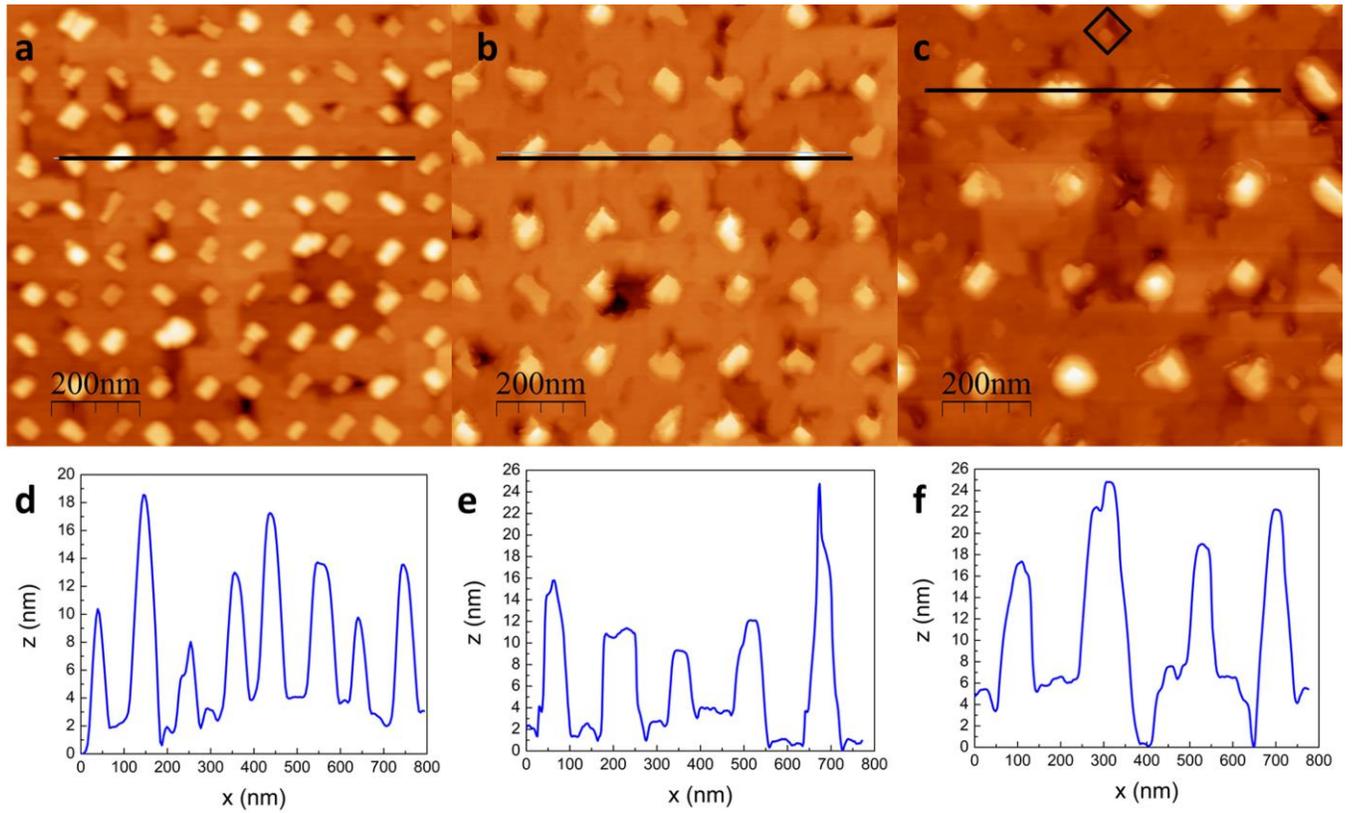

**Figure 3.** (a) Pillar topography with 100 nm center-to-center distance (pitch); (b) Pillar topography with 150 nm pitch; (c) Pillar topography with 200 nm pitch, with defect CFO pillar highlighted; (d-f) Cross section data for: (d) 100 nm, (e) 150 nm, and (f) 200 nm AFM images along the black lines shown in (a-c).

| Pillar Spacing (nm) | Mean Pillar Side Length (nm) | Mean Pillar Height Above Matrix Surface (nm) | Fractional Area Coverage (%) | CFO Pillar Volume Above Surface in Constant Area (normalized) |
|---|---|---|---|---|
| 100 | 41 | 6.1 | 16 | 1.00 |
| 150 | 55 | 7.6 | 15 | 1.02 |
| 200 | 61 | 11.8 | 9 | 1.10 |

**Table 1.** Pillar Dimensions Measured via AFM in Figure 3.

The results of the AFM and simple diffusion analysis above are a good demonstration of the nature of the templating in this work. We have shown that the CFO islands patterned on the substrate surface act as attachment sites for the epitaxial CFO pillars that form within the BFO matrix and that attachment only occurs at those sites unless there are kinetic limitations—primarily the surface diffusion length. Thus, arbitrary pitch sizes of as much as hundreds of nanometers may be possible by tailoring the growth conditions, such as the growth temperature and the deposition rate, to the template pattern. In addition, smaller pitch sizes should be readily achievable by refining the EBL and ion etching processes to improve resolution.

It is worthwhile to characterize the magnetic and ferroelectric properties of the composite as a means of comparing the template sample with other unpatterned composites in the literature. Conventional measurements to determine the magnetic anisotropy in the sample through the use of a vibrating sample magnetometer (VSM) or superconducting quantum interference device (SQUID) would be fruitless, however, as the patterned arrays cover only about 0.01% of the overall surface and any signal would be masked by the pillars that form spontaneously outside of the patterned arrays. Thus, magnetic force microscopy (MFM) is the only viable approach to study the anisotropy characteristics of the pillars.

The sample was demagnetized by an applied in-plane damped oscillatory magnetic field in order to drive the pillars to the minimum energy magnetization state. Fig. 4 shows MFM scans of the 150 nm pitch array after the demagnetization process, with a three-dimensional rendering of the topography shown and the magnetic phase overlaid as the color in the image. In Fig. 4(b)-(c),red and blue represent the positive and negative out-of-plane magnetizations (arbitrary sign), while green represents regions of small magnetic response. The nature of the magnetostatic interactions between neighboring pillars with in-plane magnetization makes the interpretation of the MFM images complicated. If the pillars are magnetized in-plane, the MFM phase contrast should be positive on one side of the pillar and negative on the other, with a neutral region in the center of the pillar due to the nature of the magnetic dipole fringe field that curls over the surface. For pillars with out of plane magnetization, the phase should be uniform across the surface of the pillar. A schematic of the two configurations is shown in Fig. 4(a).

Figure 4(d) shows the MFM phase contrast overlaid on a three-dimensional rendering of the surface topography. This technique is useful to show which pillars have in-plane and out-of-plane magnetization. Examples of magnetization along all six possible directions are highlighted in the figure. There is no clear preference for in-plane or out-of-plane magnetization, indicating that uniaxial anisotropy in the pillars is minimal and that the magnetocrystalline anisotropy in CFO along the three [100] axes dominates the magnetic behavior. This result, while different from many reports in the literature, agrees with other composite films grown by PED in our group.[27] We have previously shown[27] that the relatively slow growth rate (~1-3 Å/min) achieved using the PED technique allows for full relaxation of residual out-of-plane compressive strain in the pillars, which is the origin of the perpendicular anisotropy seen in other CFO-BFO nanocomposites.[28,29]

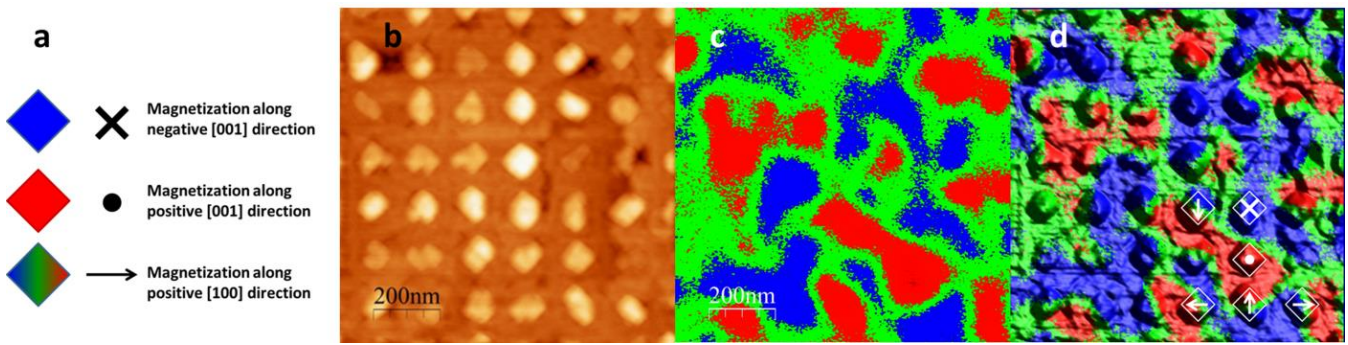

**Figure 4.** (a) Schematic of MFM phase contrasts for both in-plane and out-of-plane magnetizations; (b) Topography image of 150 nm pitch array from MFM measurement; (c) Out-of-plane MFM Phase image of array; (d) Phase contrast overlaid on three-dimensional rendering of topography with examples of magnetization along each direction shown for certain pillars.

To characterize the ferroelectric properties of the sample, piezoresponse force microscopy (PFM) was performed. PFM results topography, amplitude and phase scans are shown in Fig. 5. The amplitude scan (Fig. 5b) shows that there is no ferroelectric response from the CFO pillars, indicated by the black coloring at pillar sites. The amplitude scan also shows that there is a high density of 180° domain walls in the BFO matrix, as indicated by the dark lines throughout the pattern where the ferroelectric response vanishes.[30] The phase image shows clear domains which are 180° out-of-phase, in agreement with the amplitude scan. The domain structure of the composite is consistent with results shown elsewhere, with small 180° reverse domains forming near the interface with the pillars.[31]

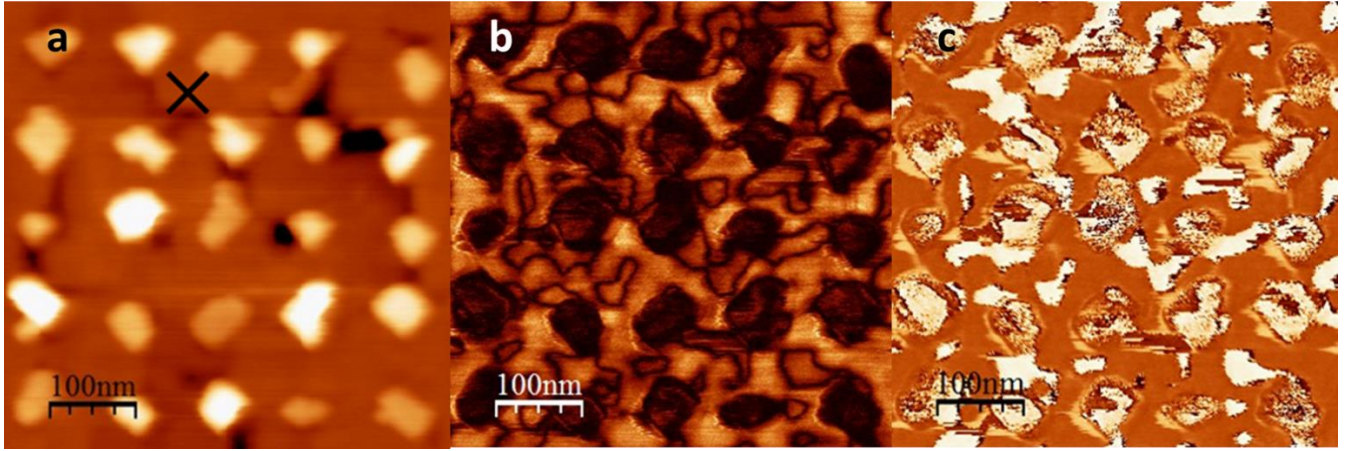

**Figure 5.** (a) Contact mode topography image taken during piezoresponse force microscopy (PFM) measurement ('X' indicates the site of switching spectroscopy-PFM (SS-PFM) measurement in Fig. 6); (b) Ferroelectric amplitude image corresponding to (a) (dark indicates low PFM response); (c) Ferroelectric phase image corresponding to (a) (white and orange correspond to oppositely oriented domains along the out-of-plane direction).

Localized ferroelectric hysteresis measurements were also performed using switching spectroscopy-PFM (SS-PFM) within the matrix region shown in Fig. 5.[32] The location of the measurement is marked with an 'X' in Fig. 5(a). Additional measurements were made at other sites in the BFO matrix within the 500 nm scan area and showed similar results. Further details of this technique and the data analysis can be found in the Supporting Information. The hysteresis loops are shown in Figure 6. Fig. 6(a) shows the out-of-plane phase, with clear switching at applied voltages of approximately 2.5 V. The measured tip displacement amplitude is shown in Fig. 6(b), with the traditional butterfly curve shape. Estimates of the value of the piezoelectric $d_{33}$ coefficient are difficult due to the complex nature of the tip-sample interactions, but estimations can be made using the measured tip displacement shown in Fig. 6b.[33] For details on this analysis, see the Supporting Information. The estimated value for $d_{33}$ is 5-10 pm/V, which is significantly lower than the 40-80 pm/V seen elsewhere in the literature for uniform BFO films.[34] However, this result is expected and has been seen in other SS-PFM measurements of nanocomposites due to the highly localized electric field originating from the small PFM tip.[35] Measurements made with electrode contacts would be expected to show somewhat larger values for $d_{33}$.

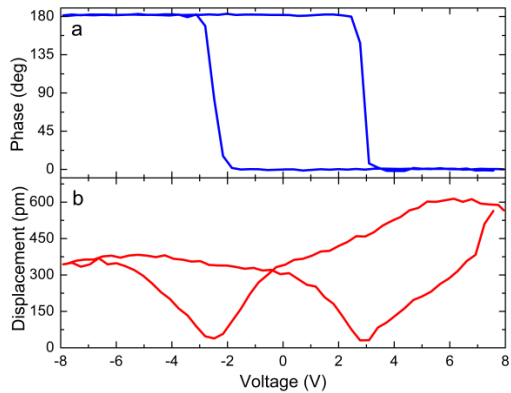

**Figure 6.** (a) Out-of-plane ferroelectric phase loop in BiFeO$_3$ (BFO) matrix measured using piezoresponse force microscopy (PFM); (b) Measured tip displacement showing the expected butterfly loop behavior..

In conclusion, we have demonstrated a new technique for the growth of ordered epitaxial CoFe$_2$O$_4$-BiFeO$_3$ nanocomposites using a top-down patterning process followed by a bottom-up template film growth. This technique is more flexible than the techniques used to create other ordered nanocomposites in that arbitrary patterns can be formed through the electron-beam lithography process. The technique also allows for the formation of the epitaxial 1-3 nanocomposites that have been shown to have large magnetoelectric response. Arrays of CFO nanopillars with pitches of 100, 150 and 200 nm have been fabricated and their structural properties have been characterized. It is shown that for both the 100 and 150 nm pitches, all CFO deposited within the patterned regions is absorbed by the template pillars, with no interstitial defect pillars forming. The magnetic properties of the CFO pillars have been characterized via MFM and the pillars have been shown to have minimal uniaxial anisotropy, which is in agreement with nanocomposite films grown without templating via PED. PFM measurements demonstrated clean ferroelectric response in the BFO matrix and hysteresis loops were measured and the $d_{33}$ out-of-plane piezoelectric response has been calculated.

ACKNOWLEDGMENT. The authors gratefully acknowledge funding from the Nanoelectronics Research Initiative, NSF (DMR-08-19762) and DARPA (HR-0011-10-1-0072). Ryan Comes also wishes to acknowledge funding from the National Defense Science and Engineering Graduate Fellowship. Research performed in part at the National Institute of Standards and Technology (NIST) Center for Nanoscale Science and Technology. The authors would also like to thank Prof. Jerrold Floro for helpful comments on the work.

SUPPORTING INFORMATION PARAGRAPH. Supporting information is available online showing the detailed growth and fabrication techniques for the initial $CoFe_2O_4$ film, amorphous Si capping layer, HSQ resist, e-beam lithography patterning and development, reactive ion etching and $CoFe_2O_4$-$BiFeO_3$ nanocomposite. Detailed explanations for the diffusion length calculations are also available along with the assumptions used. A discussion of the analysis of the piezoresponse force microscopy images and hysteresis loops is also included, detailing the calculations of the value of the piezoelectric $d_{33}$ value. This material is available free of charge via the Internet at http://pubs.acs.org.

# Supporting Information

**Sample Preparation Methodology**

*CoFe$_2$O$_4$ (CFO) Film Growth Details*

The background pressure of the Neocera PED system was approximately 5x10$^{-8}$ T and the deposition was performed in an 18 mT O$_2$ atmosphere. The substrate was held at a temperature of 600 °C during the deposition. A stoichiometric sintered CFO target was ablated using an electron pulse energy of 12.5 kV and a pulse rate of 8 Hz. The deposition rate for CFO in these conditions is approximately 0.004 Å/pulse or 1.92 Å/min. The films grow via the Volmer-Weber growth mode on the 0.5% Nb-doped SrTiO$_3$ (Nb:STO) substrate.

*Si Capping Layer Growth Details*

The Si capping layer was grown at room temperature using an RF sputter system in the NIST Center for Nanoscale Science and Technology (CNST) NanoFab. The system has a base pressure of better than 3x10$^{-6}$ T. An Ar atmosphere of 3.7 mTorr was used, with 400 W RF power applied to an undoped Si sputter target. This produces a deposition rate of 1.1 Å/sec. The Si film is expected to be amorphous and produces what appears to be a conformal coating of the CFO islands.[1]

*E-beam Lithography Patterning Details*

Dow Corning XR-1541 hydrogen silsequioxane (HSQ) negative tone e-beam resist was used for patterning. The resist was diluted in a 1:1 mixture with methyl isobutyl ketone (MIBK) to form a 3% solids content HSQ solution. The resist was spin coated on the sample at 3000 rpm for 35 seconds to produce a resist thickness of between 60 and 70 nm, as measured by optical reflectivity techniques. The electron beam lithography (EBL) was performed using a Vistec VB300 EBL system at the CNST NanoFab with area exposure doses of 5200 μC/cm$^2$ with an accelerating voltage of 100 kV. The pattern was developed using a 25% TMAH developer followed by a rinse in a 1:9 mixture of TMAH:DI water for 1 minute and then a final rinse in deionized water and blown dry using an N$_2$ gun.

*Reactive Ion Etching Details*

A Unaxis 790 RIE system was used to transfer the pattern through the Si and CFO films. An initial etch using 50 sccm $O_2$ and 5 sccm $SF_6$ was performed to selectively remove the amorphous Si capping layer. The base pressure of the system was less than $10^{-4}$ Torr and the operating pressure was 10 mTorr. The RF power was set to 100 W, which produces a DC bias of 330 V. The etch was performed for 1 minute, with the Si cap etching at approximately 7 nm/min under these conditions. The Si cap is partially removed during the process and the patterned HSQ pillars are preserved. A second etch is performed using the same RIE system, with 50 sccm Ar flow, 8 mTorr operating pressure, 500 W RF and a 750 V DC bias. This produces a CFO etch rate of approximately 6 nm/min, with Si and HSQ rates considerably higher. The etch was performed for 150 seconds to ensure that there was no residual CFO except in the patterned sites. The completion of the etch was confirmed through AFM inspection of the arrays, which showed that step-edges from the original substrate treatment were visible and that the surface topography no longer showed the island morphology that develops during the growth of the CFO film on the Nb:STO substrate.

*$CoFe_2O_4$-$BiFeO_3$ (CFO-BFO) Composite Film Growth Details*

For this deposition, the substrate temperature in the PED system was set to 615 °C and both the CFO target and a sintered $Bi_{1.2}FeO_3$ ($B_{1.2}FO$) were used. A 15 mT $O_2$ atmosphere was used during the deposition. Each target was ablated using a separate electron gun, with both guns set to 12.5 kV pulse energy. The $B_{1.2}FO$ target was pulsed at a rate of 5 Hz and the CFO target was pulsed at a rate of 1 Hz, which produces composite films with between 10 and 20% CFO by volume when grown on an unpatterned substrate. The deposition rate for both BFO and CFO in these conditions is approximately 0.005 Å/pulse.

**Calculation of CFO Adatom Diffusion Length**

The work of Zheng *et al*. (Ref. 2) estimated the activation energy for surface diffusion of CFO adatoms during PLD growth.[2] In this work, the authors grew a series of unpatterned BFO-CFO nanocomposites at various deposition rates and temperatures using a single 1:1 CFO:BFO composite target. They determined the lateral size of the resulting CFO pillars, *d*. By assuming that the diffusion length was approximately equal to *d*, they showed that the diffusion length was well modeled by assuming:

$$d = \sqrt{4Dt} \qquad (1),$$

where $D$ is the surface diffusivity and $t$ is a characteristic time inversely proportional to the deposition rate. They were able to show that the diffusivity, $D$, had the expected Arrhenius temperature dependence:

$$D = D_0 e^{-E_a/kT} \qquad (2),$$

where $E_a$ is the activation energy for diffusion, $k$ is Boltzmann's constant, $T$ is the growth temperature and $D_0$ is a constant that defines the diffusivity in the limit $kT \gg E_a$. By fitting the temperature dependence of $d$ they were able to estimate the value of $E_a$ as 1.66 eV for CFO adatoms on the surface of the BFO matrix. No uncertainty in $E_a$ was published, but based on comments in the text of the article stating that the value is reasonable when compared with that for CFO adatoms in CFO-BaTiO$_3$ (BTO) composites (1.56 eV), we have assumed an uncertainty of ±0.05 eV for $E_a$ in our calculations. No mention is made of the value for $D$ or $D_0$, which is critical for a more precise estimation of the diffusion length.

In order to determine $D_0$, the data from the paper was tabulated and fits were performed.[2] A table showing the pillar diameter, $d$, and the growth rate, $v$, from the paper is shown in Table 1. In the original work, the authors assumed that the diffusion length was approximately equal to the pillar diameter. However, a more precise estimation is required to quantify $D_0$. Since the composition of the films was 1:1 BFO:CFO, it is possible to geometrically determine the maximum distance that an adatom must travel if we assume that the pillars are arranged in a close-packed (hexagonal) array. While this assumption is more valid for CFO-BTO composites than it is for CFO-BFO composites, it should still improve the estimate when compared to the assumption presented in the paper.[3] The maximum capture radius, $r$, for a pillar in a close-packed array was determined geometrically for a 50% pillar area coverage and found to be:

$$r = d / \sqrt{3\sqrt{3}} \qquad (3).$$

The relevant time, $t$, was calculated by assuming that adatoms would only diffuse for a period of time equal to the time required to form one unit cell monolayer of BFO (~4.1 Å if BFO is strained by the STO substrate[4]) in the matrix. The capture radius and monolayer formation time are also shown in Table 1. A linear fit was performed to determine the slope of a line:

$$r = C\sqrt{t} \tag{4},$$

where C is an unknown constant that depends on $D_0$. From this fit, $D_0$ may be extracted using:

$$D_0 = \frac{C^2}{4} e^{E_a/kT} \tag{5}.$$

The value of $D_0$ was found to be $1.2\pm0.8\times10^{10}$ nm$^2$/sec after accounting for the uncertainty in both the fit and the assumed uncertainty in $E_a$ from the published work.

| Pillar diameter, $d$ (nm) | Capture radius, $r$ (nm) | Growth rate, v (nm/min) | Monolayer formation time, $t$ (sec) |
|---|---|---|---|
| 235 | 103 | 0.5 | 49 |
| 170 | 77 | 2 | 12 |
| 125 | 55 | 4 | 6 |
| 100 | 44 | 8 | 3 |

Table 1: Published data for CFO pillar diameter and film growth rate taken from Zheng, *et al.* (Ref. 2).[2] Calculated data for capture radius and monolayer formation time.

With this result, it is possible to calculate the diffusion length of the CFO adatoms in this work. Using Equation (1) and assuming a reasonable uncertainty in substrate temperature of 25 K based on the nature of our PED heating system, we find that the diffusion length of the CFO adatoms is 55±26 nm. Since the capture radius for a pillar in an array with 150 nm pitch is 79 nm, the calculations demonstrate that the diffusion length may be long enough for all adatoms to diffuse to the template sites. However, for the 200 nm pitch the capture radius of 111 nm is not in agreement with the uncertainty of the diffusion length.

**Sample Characterization Techniques**

*X-ray Reflectivity*

The sample was characterized using x-ray reflectivity in a Rigaku Smart-Lab diffractometer to measure the film thickness in the sample. All oscillations were assumed to originate from the BFO matrix given the relatively small area fraction of CFO covering the surface and the facetted structure of the CFO pillars that form.

The data was well modeled by this technique, with a density of 7.5 g/cm$^3$, a film thickness of 25 nm and a root-mean-square surface roughness of 1.2 nm. The calculated density is slightly less than that of bulk BFO (8.3 g/cm$^3$)[5], which can be attributed to the presence of the CFO pillars since CFO has a bulk density of 5.2 g/cm$^3$.[6] The CFO pillars can likewise explain the relatively high surface roughness measured.

*Piezoresponse Force Microscopy*

Piezoresponse force microscopy (PFM) was performed using an Asylum Research Cypher AFM system. The measurements were performed using the Dual AC Resonance Tracking (DART) technique.[7,8] The DART technique measures the ferroelectric response near the contact resonance of the cantilever, allowing for enhanced signal-to-noise ratio and better contrast in the PFM amplitude and phase images. The drawback for these measurements is that rough features such as the non-ferroelectric CFO pillars will become noisy in the phase image. This can be seen in Fig. 5(c) of the main article. However, the smooth BFO matrix is well measured and shows good amplitude and phase contrast. The validity of this technique can be confirmed by noting that the PFM amplitude image in Fig. 5(b) (in the main article) shows that there is no ferroelectric response at the locations of the CFO pillars.

An Asylum Research ASYELEC-01 Ir coated tip was used for the measurements, with the conductive Nb-doped SrTiO$_3$ serving as a bottom electrode. The tip has a free air resonant frequency of 86.1 kHz and a contact resonant frequency, $f$, of 310.9 kHz and a quality factor, $Q$, of 132. Switching-spectroscopy PFM (SS-PFM) hysteresis measurements were performed using the tip to switch small domains in the matrix. These results are shown in Fig. 6 of the main article and are reproduced below for convenience as Fig. 1. The voltage was scanned three times over a range of +8 V to -8 V applied tip bias and the ferroelectric response was measured each time at 0 V applied bias, with an alternating current drive amplitude, $V_{ac}$, of 500 mV. This technique is standard for SS-PFM measurements.[9] The resulting PFM amplitude and out-of-plane phase loops were then averaged to produce the loops shown in Fig. 1(a) and 1(b). The value of the $d_{33}$ coefficient is proportional to the measured PFM tip displacement, $\Delta z$, according to the formula:

$$d_{33} = \Delta z / Q * V_{ac} \qquad (6).^8$$

The maximum values for $d_{33}$ occur in the range of ±4-8 V DC bias and have maximum values of ~5.5 pm/V for negative biases and ~9 pm/V for positive biases. The value for $d_{33}$ is thus roughly estimated to be 5-10 pm/V, which is presented in the main article.

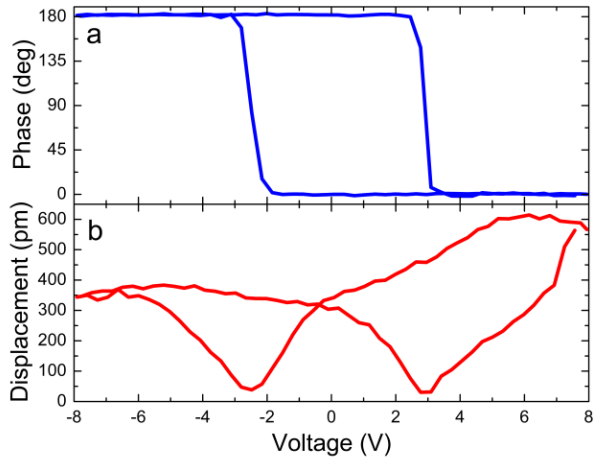

**Figure 1- (a) Phase response; (b) Measured tip displacement.**